\documentclass{ws-ijmpa}

\begin{document}

\markboth{Kuzmichev}{Tunka project}

\catchline{}{}{}{}{}

\title{The Tunka Experiment: Towards a 1-km$^2$ Cherenkov EAS Array in the Tunka
Valley.}

\author{D.Chernov, N.Kalmykov, E.Korosteleva, L.Kuzmichev, V.Prosin, M.Panasyuk,
A.Shirokov, I.Yashin}
\address{Scobeltsyn Institut of Nuclear Physics of MSU, Moscow, Russia}

\author{N.Budnev, O.Gress, L.Pankov, Yu.Parfenov, Yu.Semeney}
\address{Institute of Applied Physics of ISU, Irkutsk, Russia}  

\author{B.Lubsandorzhiev, P.Pohil}
\address{Institute for Nuclear Research of RAS, Moscow, Russia}  

\author{V.Ptuskin}
\address{IZMIRAN, Moscow, Russia}

\author{Ch.Spiering, R.Wischnewski}
\address{DESY-Zeuthen, Zeuthen, Germany}  

\author{G.Navarra}
\address{ Universita' Torino, Italy}


\maketitle


\begin{abstract}

The project of an EAS Cherenkov array in the Tunka valley/Siberia 
with an area of
about 1 km$^2$ is presented. The new array will have a ten times bigger area
than the existing Tunka-25 array and will permit a detailed study of the
cosmic ray
energy spectrum and the mass composition in the energy range from 10$^{15}$ to
10$^{18}$ eV.

\end{abstract}



\section*{Introduction}

The nature of Galactic sources of high and ultrahigh energy cosmic rays is not
clear, despite of essential progress in the theory of their acceleration and
propagation. To understand the nature of
sources, reliable experimental data about energy spectrum and mass
composition in the range of 10$^{15}$ - 10$^{18}$ would be essential.
This range includes as well the  "classical" knee [1] at energy $3\cdot 10^{15}$ eV,
as also the range of energies $10^{17} - 10^{18}$ eV, considered now as maximum
reachable by cosmic rays accelerated in Galactic supernova remnants 
[2]. 

It is known, that the most reliable method of cosmic rays energy measurement in 
the specified energy range is the method of EAS Cherenkov light observation.
  
A series of such experiments are carried out in Tunka Valley 
(50 km from lake Baikal) since
more than 10 years: Tunka-4 [3], Tunka-13 [4] and Tunka-25 [5]. The latter array 
consists of 25 detectors on the basis of PMT QUASAR-370 and 4 detectors
with pulse shape recording on the basis of PMT Torn-EMI D669. 
To study charged particles at surface,
water Cherenkov detectors with an area of 10 m$^2$ are planned to be added to  
the array. The detectors cover an area of $\sim 0.1$ km$^2$. 

The construction of the TUNKA-133 array will allow us to expand 
the sensitive area by almost
10 times. For one season of data taking (400 hours) we plan to collect $\sim 3\cdot
10^5$ events with energies above 3$\cdot 10^{15}$ eV, $\sim 200$ events with
energies higher than $10^{17}$ eV, and a handful events with energies higher than      
$10^{18}$ eV.

\begin{figure}
\centerline{\psfig{file=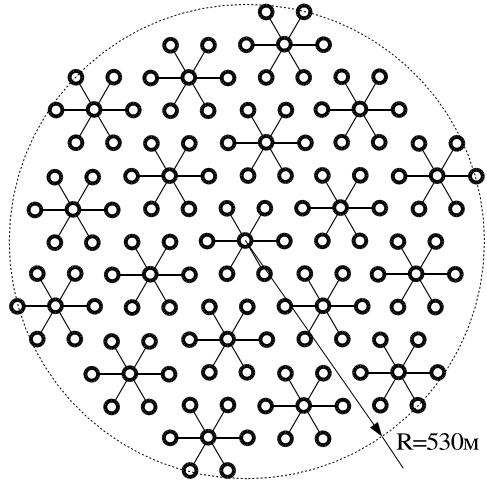,width=0.5\textwidth}}
  \caption{\textit{TUNKA-133 array. Minimal distance between
detectors is 85 m. Seven detectors form a cluster.}}
\end{figure}

\section*{Array Description}
The new array will consist of 133 optical detectors on the basis of PMT
EMI 9350 (diameter of photocathode of 20 cm). 
The plan of the array is shown in fig. 1. The 133 detectors
are grouped in 19 clusters of seven detectors each.
The analog electronics is designed to provide a high dynamic range (10$^4$). 

The functional scheme of the cluster electronics is shown in fig. 2. The 
electronics will be placed inside industrial computers with ISA interface.
Data acquisiton and management will be carried out by
a single-board industrial computer PCA-6145, connected with the control center 
through Ethernet. 

The measuring channels will be designed on the basis of 10-bit ADCs AD9410
and Xilinx microchip FPGAs. The continuous digitizing of  signals with step
widths 
of 5 ns allows to discard traditional TDCs and to replace analog discriminators
by digital ones. Digitizing of signals at the central detector of each cluster,
with steps of 2.5 ns, allows to measure the pulse form and to deduce the FWHM.

The cluster trigger condition will be $\ge 3$  detectors during the
resolution time of 0.5 microsec. The rate of single-cluster triggers 
will be about 2 Hz, and the cluster data stream about $\sim$2 Kb/sec.

\begin{figure}
\centerline{\psfig{file=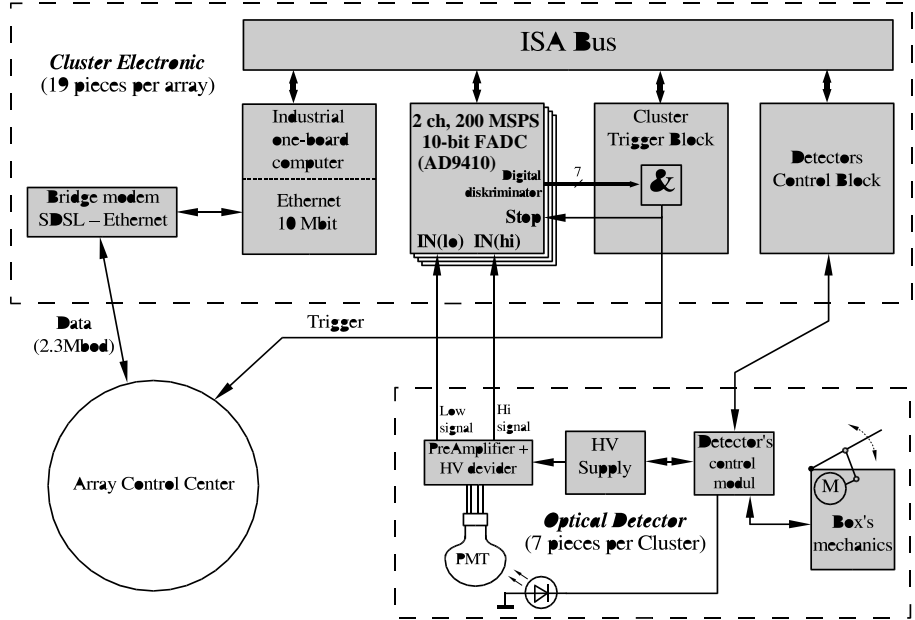,width=0.95\textwidth}}
  \caption{\textit{Functional scheme of cluster electronics.}}
\end{figure}

The reconstruction of EAS parameters will be carried out with methods, developed
for the TUNKA-25 array [6]. These methods provide an accuracy of core location of
about 6 m and a primary energy resolution of about 15\%. The error in $X_{max}$ 
will be about 35 g/cm$^2$
for the LDF-steepness method, and about 25 g/cm$^2$ for the pulse
FWHM method.


\section*{Acknowledgements}

Present work is supported by Russian Fund of Basic Researchs
(grants 02-02-17162 and 03-02-16660). The PMTs EMI 9350 (200) have
been provided by INFN from the former MACRO experiment.


\end{document}